\documentclass{elsart}

\usepackage{epsfig}

\usepackage{amssymb}

\begin{document}

\begin{frontmatter}



\title{Continuum percolation of \\ 
       wireless ad hoc communication networks}

\author[label1,label2]{Ingmar Glauche}
  \ead{glauche@physik.phy.tu-dresden.de}
\author[label1,label3]{Wolfram Krause}
  \ead{krause@th.physik.uni-frankfurt.de}
\author[label1]{Rudolf Sollacher}
  \ead{rudolf.sollacher@mchp.siemens.de}
\author[label1]{Martin Greiner}
  \ead{martin.greiner@mchp.siemens.de}
\address[label1]{Corporate Technology, Information \& Communications, 
                 Siemens AG, 
                 D-81730 M\"unchen, Germany}
\address[label2]{Institut f\"ur Theoretische Physik,
                 Technische Universit\"at,
                 D-01062 Dresden, Germany}
\address[label3]{Institut f\"ur Theoretische Physik,
                 Johann Wolfgang Goethe-Universit\"at,
                 Postfach 11 19 32,
                 D-60054 Frankfurt am Main, Germany}

%

\begin{abstract}
Wireless multi-hop ad hoc communication networks represent an
infrastructure-less and self-organized generalization of todays
wireless cellular networks. Connectivity within such a network is an
important issue. Continuum percolation and technology-driven mutations
thereof allow to address this issue in the static limit and to
construct a simple distributed protocol, guaranteeing strong
connectivity almost surely and independently of various typical
uncorrelated and correlated random spatial patterns of participating
ad hoc nodes.
\end{abstract}

\end{frontmatter}

\newpage
\section{Introduction}
\label{sec:intro}

Today's wireless communication mainly relies on cellular networks
\cite{PRO95,PRA98,RAP99}. At first, the sending mobile device directly 
connects to its nearest base station. A backbone network then routes
the communication packets to the cell, where the intended receiving
mobile device is registered. Finally, the cell's base station
transmits the passed-by message to the latter. As part of the
centralized backbone infrastructure each base station acts as a
router, possesses the network information, controls the single-hop
communications within its cell and assigns different channels to its
various mobile clients. The base stations need to be placed according
to some optimized coverage layout. This requires an enormous planning
effort ahead of operation and leads to a static
infrastructure, hard to change and adapt to new, revised
needs. This costly inflexibility motivates a flexible and
infrastructure-less peer-to-peer concept: selforganizing wireless
mobile ad hoc networks \cite{HAA99,HUB01,MANET,NISTB}.

In a wireless ad hoc network, a sending mobile device uses inbetween
mobile devices to communicate with the intended receiver. Such a
multi-hop connection requires each mobile device to have additional
router functionality. As a central control authority is missing, the
participating devices need coordination amongst themselves to ensure
network connectivity, efficient discovery and execution of end-to-end
routes and avoidance of data packet collisions on shared radio
channels; of course, mobility of the devices also has an impact on the
network performance, which has to be coped with. Contrary to these
global network features, the selforganizing coordination rules, called
protocols in the jargon of electrical engineers, have to be local. Due
to its limited transmission range, a mobile device is able to
communicate only with its current spatial neighbors. Hence, it can
only extract information on its local surrounding. Since this is the
only input into the coordination rule, the latter is by definition
local. Upon execution, it readjusts for example the device's
transmission power to its new surrounding.

In this Paper we focus on the important connectivity issue and ask:
what is a good local coordination rule for transmission power
management, which almost surely guarantees global connectivity for the
whole network?  We employ a simple static model for ad hoc
communication networks.  This allows a connection to continuum
percolation theory \cite{MEE96,STO95} based on random geometric graphs
\cite{DAL02,BAK02}. The spatially distributed ad hoc devices
correspond to nodes, which are more or less locally connected by
communication links. Two nodes establish a mutual link, only if the
first node lies within the transmission range of the second and vice
versa. For the case of constant, isotropic transmission ranges, a
further mapping onto the classical picture of continuum percolation
\cite{ISI92,BLA02}, stemming from the transport physics in continuous
random media, is straightforward: whenever discs with radius equaling
half of the transmission range are placed around two nodes and
overlap, the two nodes are linked.

Continuum percolation allows to study, for example, the dependence of
the probability for strong connectivity on the transmission range and
to find a critical range, above which the ad hoc network graph is
almost surely connected. The critical transmission range can be
translated into a critical node neighborhood degree $ngb_{\rm
crit}$. Hence, a simple local coordination rule would be for each node
to adjust its transmission power to yield a little above $ngb_{\rm
crit}$ neighbors. As we will demonstrate, such a rule is not flexible
enough to perform equally well in various different environments, like
homogeneous vs.\ heterogeneous random spatial arrangements of nodes or
homogeneous vs.\ heterogeneous propagation media. Fortunately, such a
rule is able to give some guidance to develop local coordination rules
with improved adaptation properties.  In some respect, these new rules
represent technology-driven mutations of the continuum percolation
problem and demonstrate the usefulness to combine real-world needs in
electrical engineering with modified concepts of statistical physics.

In Section 2 a precise definition of random geometric graphs in the
context of wireless ad hoc communication networks is given; various
spatial point patterns, uncorrelated and correlated, are
introduced. In Section 3 standard continuum percolation based on discs
with constant and random radius is applied to ad hoc networks to
obtain first rigorous statements on coordination rules for
connectivity. A mutation of continuum percolation is presented in
Section 4, which directly leads to a distributed, local coordination
rule, flexible enough to cope with various environments. A conclusion
and outlook is given in Section 5.

\section{Random geometric graph approach to ad hoc networks}
\label{sec:two}

The topology of static ad hoc networks can be viewed as a random
geometric graph. It consists of a spatial pattern of points, where
each point is connected to some others by links. In Subsect.\ 2.1 the
construction of three generic random point patterns is described. A
simple propagation-receiver model is used in Subsect.\ 2.2 to
establish links between points. Some selected geometric-graph features
of interest are discussed in Subsect.\ 2.3.

\subsection{Spatial point patterns}

Throughout this Paper three different generic random spatial point
patterns are used: a homogeneous, a multifractal and a Manhattan point
pattern. In the following comes a short description of their
construction. With no loss of generality a two-dimensional square area
with side length $L=1$ is used.

In a random homogeneous point pattern each of the $N$ points is given
a random position $(x,y)\in[0,L]\times[0,L]$. A typical realization is
illustrated in Fig.\ 1a. By definition it does not show generic
clustering.

One way to construct simple clustered point patterns is to employ a
binary multiplicative branching process. The nonuniform probability
measure supported on the unit square is constructed by iteration: at
first the parent square is divided into four offspring squares with
area $1/4$. Two randomly chosen offsprings get a fraction
$(1+\beta)/4$ of the parent probability mass $\mu=1$, whereas the
remaining two get a fraction $(1-\beta)/4$. In the next iteration step
each offspring square follows the same probabilistic branching rule
and nonuniformly redistributes its probability mass onto its own four
offsprings.  After $j$ iteration steps the probability mass $\mu=1$
has been nonuniformly subdivided onto $4^j$ subsquares with area
$1/4^j$, where ${j \choose i} 2^j$ of these subsquares $(0\leq i \leq
j)$ come with probability mass $[(1+\beta)/4]^i
[(1-\beta)/4]^{j-i}$. One after the other each of the $N$ points to be
distributed is given an independent and uniform random number between
$0$ and $1$, which, given some probability-mass-weighted ordering of
the $4^j$ subsquares, corresponds to exactly one subsquare, onto the
particle is deposited and randomly placed inside. One such realization
of a point pattern is shown in Fig.\ 1b. The hierarchical clustering
of points is due to the hierarchical branching structure of the
iteration process.  -- The probability measure constructed with such a
multiplicative branching process is a multifractal \cite{FED88}. The
construction of multifractal fields has some importance in such
diverse fields as turbulence
\cite{FRI95,MEN91,GRE96,GRE98}, finance \cite{LUX01,MUZ00}, Internet 
traffic \cite{RIB01}, high-energetic multiparticle dynamics
\cite{WOL96} and deterministic chaos \cite{HAL86}, just to name a few.

As a third generic class of spatial point patterns a Manhattan street
pattern is used. $N_x$ and $N_y$ streets are equidistantly placed
parallel to the x- and y-axis, respectively. One after the other each
of the $N$ points is randomly placed onto one randomly chosen
street. Fig.\ 1c gives an illustration of one realization.

\subsection{Construction of communication links}

For a wireless communication network we define a link between two
nodes $i$ and $j$, only if they can communicate back and forth to each
other.  Let $P=P_i$ denote the transmission power given to node $i$,
then according to a simple propagation-receiver model, which does not
account for shadowing and fast-fading effects, node $j$ is able to
receive the signal once
\begin{equation}
\label{eq:zweib1}
  \frac{ P/r^\alpha }{noise}  \geq  snr \; .
\end{equation}
$r=r_{ij}$ is the relative Euclidean distance between nodes $i$ and
$j$.  The path-loss exponent $\alpha$ is assumed to be constant; for
free-space propagation it is $\alpha=2$, but depending on specific
in-/outdoor propagation it can vary typically between $1 \leq \alpha
\leq 6$. For a successful signal transmission the received power
$P_i/r_{ij}^\alpha$ relative to a noise power needs to be larger than
the minimum signal-to-noise ratio $snr$. Without any loss of
generality the variables $noise$ and $snr$ can be set equal to one,
implying a rescaling of the power $P_i$. Condition (\ref{eq:zweib1})
guarantees that node $j$ is able to hear node $i$. This alone would
define a directed link $i \to j$.  Since bidirectionality increases
the efficiency of the communication feedback, we focus on
bidirectional links ($i \leftrightarrow j$): also $i$ needs to hear
$j$, implying that also (\ref{eq:zweib1}) has to be fulfilled with the
substitution $P_i \to P_j$. We call a bidirectional link a
communication link.

The link construction can be given a simple geometric interpretation.
With $P=P_i$ the condition (\ref{eq:zweib1}) translates into a maximum
range $R_i=P_i^{1/\alpha}$. Nodes that lie inside this circle with
radius $R_i$ around node $i$ are able to hear this node. For a
communication link to exist between nodes $i$ and $j$, $j$ has to lie
inside $i$'s circle with radius $R_i$ and $i$ has to lie inside $j$'s
circle with radius $R_j$. For the case that all $R_i=R$ are identical,
this link construction matches the standard link construction of
continuous percolation \cite{ISI92}.

Besides various spatial point patterns Fig.\ 1 also illustrates the
communication links. Each node has been given identical power
$P/P_{\rm norm}=5$, where the normalization
$P_{\rm norm}=R_{\rm norm}^\alpha$ 
comes by setting ${\pi}R_{\rm norm}^2=L^2/N=1/\rho$
equal to the reciprocal of the node density $\rho$. Note that no
periodic boundary conditions have been used for this figure.

\subsection{Geometric-graph features of interest}

In Sections 3 and 4 various rules for assigning power values to the ad
hoc nodes will be discussed. Each such rule together with a chosen
generic class of spatial point patterns defines a specific ensemble of
geometric graphs. For example, the graphs of Figs.\ 1a-c each
represent one realization out of three different ensembles: (a)
identical power for a random homogeneous spatial point pattern, (b)
identical power for a multifractally clustered random spatial point
pattern, and (c) identical power for a Manhattan random spatial point
pattern.

One question to ask for each ensemble of geometric graphs is for
example: how large is the average giant component? The giant component
is defined as the magnitude of the largest connected cluster appearing
in a graph realization; see again Fig.\ 1 for a visualization. A
related and important question to ask for ad hoc communication
networks is, what is the probability that all nodes are able to
communicate to each other via multihop link routes? With other words,
what is the probability that the giant component is equal to the
number of nodes $N$? As connectivity is for sure a very important
issue for ad hoc communication networks, so is a generalization called
$k$-connectivity. A graph is called $k$-connected if between every
pair of nodes there exist at least $k$ independent paths, which
implies, that once $k-1$ nodes are removed at random, the graph
remains at least one-connected. A $k$-connected ad hoc communication
network is more flexible and robust to routing failure.  Hence,
another question: what is the probability for an ad hoc geometric
graph to be $k$-connected?

A simple flooding algorithm is used to determine the giant component
of an ad hoc graph realization: a random node is tagged in first
place, then its neighbors are tagged, which then continue to tag their
untagged neighbors, and so on, until the corresponding cluster is
saturated. This procedure is repeated for all untagged nodes, until
all nodes of the graph are tagged. By definition, the largest found
cluster is equal to the giant component. For an ad hoc geometric graph
to be one-connected, i.e.\ strongly connected, the giant component has
to equal the total number of nodes. Another procedure to inquire
one-connectivity uses the $N{\times}N$ Laplace matrix \cite{BOL85},
which is the difference between the diagonal node degree matrix and
the adjacency matrix. An element of the adjacency matrix is either one
or zero, depending on whether a link does or does not exist between
the two respective nodes; a diagonal element of the node degree matrix
counts the link neighbors of the respective node. If the Laplace
matrix possesses only one zero eigenvalue, then the graph is
one-connected; the number of zero eigenvalues counts the number of
partitioned clusters. We have employed the Laplace matrix algorithm
only as a supplement for small $N$ geometric graphs. The probability
for one-connectivity is estimated from a representative sample of ad
hoc geometric graph realizations as the ratio between one-connected
and all sample graphs.

($k{>}1$)-connectivity is algorithmically very costly. In a nutshell,
for each of the $N(N-1)/2$ pair of nodes belonging to one graph it has
to be checked whether at least $k$ independent paths exist. Although
some faster, but approximate algorithms can be found in the recent
literature \cite{WHI01}, we prefer to switch from costly $k$- to cheap
pseudo-$k$-connectivity. Given one-connectivity, the latter only
requires each node to have at least $k$ neighbors. In fact, a theorem
exists \cite{PEN99}, which, translated into our language, guarantees
for a geometric graph ensemble based on random homogeneous point
patterns and the artificial constant-power link rule of Sect.\ 3.1
that in the large $N$ limit the probabilities for $k$- and
pseudo-$k$-connectivity converge as they approach one.

\section{Continuum percolation with artificial link rules}
\label{sec:three}

\subsection{Artificial link rule I: constant transmission power}

The simplest rule for power assignment is to allocate the same power
value $P_i=P$ to each node $i$. This is an artificial and unrealistic
rule. It would require all nodes either to be designed for only the
same single-valued power operation or, given already network
connectivity, to carry out fast synchronization; of course, also an
outside provider could adjust all node powers to one single value, but
this would give up the infrastructureless philosophy of ad hoc
networks. Nevertheless, the constant-$P$ rule is good to start with
\cite{GUP00}.

Fig.\ 2 shows the average relative giant component as a function of
$P$ obtained from a sample of $500$ geometric graphs generated with
random homogeneous spatial point patterns and the constant-$P$ rule.
The chosen rather small sample size produces already more than
sufficient statistical convergence and keeps statistical error bars to
such small values, that those will not be shown for this and the
following figures.  A percolation threshold behavior around 
$P_{\rm crit} \approx 4.5 P_{\rm norm}$ is observed: for 
$P \ll P_{\rm crit}$ the average relative giant component is close to 
zero, whereas for $P\gg P_{\rm crit}$ it is almost equal to one.  The 
sharpness of the threshold depends on the number of participating 
nodes; with increasing $N$ the transition becomes sharper.

For the determination of the exact threshold position the limit
$N\to\infty$ would be needed. To partially account for this, the
relative giant component has also been determined from small and
medium $N$ simulations with periodic boundary conditions. Results are
also shown in Fig.\ 2. The percolation transition becomes sharper and
moves a little to the left once compared with the previous results,
but still no full convergence for the employed increasing $N$ values
is obtained. Since for realistic ad hoc networks the limit
$N\to\infty$ is out of reach, we do not pursue this matter
further. For the rest of this Paper we focus on $N=1600$, which
accommodates best a hot-spot, i.e.\ big-crowd application of wireless
ad hoc networks.

The critical power $P_{\rm crit} \approx 4.5 P_{\rm norm}$, obtained
by setting the path-loss exponent equal to $\alpha=2$, can be given a
more illustrative interpretation: the factor $4.5$ reflects the
average number of neighboring nodes, which is also denoted as the
average node degree.  Defining $P_{\rm crit} = R_{\rm crit}^2$ along
the lines of relation (\ref{eq:zweib1}), we get 
$\langle k \rangle = {\pi}R_{\rm crit}^2\rho 
\approx 4.5\pi R_{\rm norm}^2\rho = 4.5$.
Picking one node, the probability to find $k$ other nodes inside its
circular disc with radius $R_{\rm crit}$ is equal to 
$p(k) = {N-1 \choose k} q^k (1-q)^{N-1-k} 
\approx (\lambda^k/{k!}) e^{-\lambda}$,
where $q={\pi}R_{\rm crit}^2/L^2$. For $N$ large and $q$ small, $p(k)$
becomes a Poissonian with mean 
$\langle k \rangle = \lambda = q(N-1)\approx qN \approx 4.5$.  Note, 
that due to finite-size effects and the usage of no periodic boundary 
conditions the actually sampled critical link degree is a little 
smaller than the asymptotic value $k_{\rm crit}=4.53$, which is stated 
for example in \cite{DAL02}.
{--} 
This demonstrates that other than in terms of $P_{\rm crit}$ it
is also convenient to characterize the percolation phase transition in
terms of $R_{\rm crit}$ or $k_{\rm crit}$. Once choosing other
path-loss exponents $\alpha \neq 2$, the former will change according
to 
$P_{\rm crit}(\alpha{\neq}2) = 
R_{\rm crit}^{\alpha-2} P_{\rm crit}(\alpha{=}2)$, whereas 
$R_{\rm crit}$ or $k_{\rm crit}$ remain as before.

Next, spatial point patterns other than random homogeneous are
discussed.  Fig.\ 3 compares the relative average giant component as a
function of transmission power obtained for random multifractal and
Manhattan spatial point patterns with the random homogeneous case;
consult again Fig.\ 1.

Evidently for small $P$ the relative giant component is larger for the
multifractal than for the homogeneous patterns, but for the
convergence of the relative giant component towards one a much larger
$P$ is needed. Due to the hierarchical clustering, subclusters of
points are easily formed at small $P$ since only a small transmission
range is needed to connect the corresponding nodes. However, in order
to connect the various subclusters either directly to each other or
via isolated nodes lying inbetween it needs a rather large
power. Another consequence of the pronounced clustering is that the
rather sharp percolation threshold observed for homogeneous point
patterns blurs more the larger the splitting parameter $\beta$
characterizing the multifractal point patterns is chosen.

Also for the Manhattan point patterns the relative giant component
increases faster for small transmission powers than for the
homogeneous point patterns. The reason is that the average
nearest-point distance is smaller for the random points confined to
the one-dimensional Manhattan streets. The Manhattan threshold for the
relative giant component is relatively sharp and comes at a
transmission power value, which is somewhat smaller than for random
homogeneous point patterns.
{--} 
For consistency, it is instructive to map the continuous
percolation based on random Manhattan point patterns onto the
well-known square-lattice bond percolation \cite{STA92}. Two nodes are
assumed to lie on a one-dimensional straight line and to have a
distance $l=L/N_x=L/N_y$ corresponding to the distance of two
successive Manhattan street crossings. They can only communicate with
each other, if inbetween nodes come with successive distances $x$
smaller than their transmission range $R$; otherwise the occurring
void is too large to be bridged. For a one-dimensional Poissonian
point pattern with density $\lambda=N/(N_x+N_y)$ the distance $x$ of
two consecutive nodes is exponentially distributed according to
$p(x)=\lambda\exp{(-\lambda x)}$. This allows to estimate the
probability for the occurrence of at least one too large void between
the two picked nodes with distance $l$. The expression is
\begin{equation}
\label{eq:dreia1}
  p({\rm void}>R)
    =  \sum_{j=1}^{m} \frac{(-\lambda)^{j-1}}{(j{-}1)!} 
       (l-j R)^{j-1} \left[ 1 + \frac{\lambda}{j}(l-jR) \right]
       e^{-j\lambda R}
       \; ,
\end{equation}
where $m=\lfloor l/R \rfloor$; a derivation of this formula is given
for example in Ref.\ \cite{DOU02}. Setting the path-loss exponent to
$\alpha=2$ and according to (\ref{eq:zweib1}) converting the threshold
power $P \approx 4 P_{\rm norm}$ into $R$ we arrive at a value 
$p({\rm void}{>}R)=0.47$, which almost matches the critical bond 
probability $p_{\rm bond}\!=0.50$ of bond percolation on a square 
lattice \cite{STA92}.

As long as the average relative giant component is not exactly one, no
direct knowledge on the probability for strong connectivity, i.e.\
one-connectivity, is possible. The ($k{=}1$) curve of Fig.\ 4
illustrates this quantity as a function of $P$ obtained from a sample
of $500$ geometric graphs generated with random homogeneous spatial
point patterns and the constant-$P$ rule. It also reveals a threshold
behavior, which sets in once the relative giant component has
approached one. The relative factor of the respective threshold
positions is about $2.3$. The probabilities for pseudo-$k$
connectivity with $k\!=\!2$ and $3$ are also depicted in Fig.\ 4. Of
course, their threshold is shifted to even larger $P$ when compared to
the pseudo-one threshold.  
-- 
A part of Fig.\ 5 shows the probability for strong connectivity for
random multifractal and Manhattan point patterns. Whereas the
Manhattan curve is close to the homogeneous curve, the onset for a
nonvanishing probability in case of the multifractal point patterns is
shifted to extremely large values of the transmission power. This is
due to the inhomogeneous spatial clustering and demonstrates that the
constant-$P$ rule is not efficient in such an environment.

\subsection{Artificial link rule II: iid transmission power}

So far all nodes were assigned the same transmission power $P$. This
constant-$P$ rule is not able to counterbalance spatially sparse
regions, which remain disconnected to the giant component. Some
long-range links are called for to establish connections between
otherwise separated subclusters. Some ad hoc nodes then have to send
with a transmission power larger than average. The simplest
heterogeneous rule in this context is the iid-$P$ link rule
\cite{MAR02}. It treats the transmission power values assigned to the
ad hoc nodes as independently and identically distributed (iid) random
variables. Independently from the other nodes each node chooses its
power according to the same probability distribution $p(P)$ with mean
$\langle{P}\rangle$. Translating $P$ into a transmission range $R$,
this rule places a disc with random radius $R$ around each node.

As a flexible representative a bimodal distribution 
\begin{equation}
\label{eq:dreib1}
  p(P) = \frac{\beta_2}{\beta_1+\beta_2}
         \delta\Bigl( P-\langle{P}\rangle(1-\beta_1) \Bigr)
         + \frac{\beta_1}{\beta_1+\beta_2}
         \delta\Bigl( P-\langle{P}\rangle(1+\beta_2) \Bigr)
\end{equation}
is chosen. It comes with two parameters $\beta_1$ and $\beta_2$
determining the variance and skewness of this distribution, but
leaving the mean $\langle{P}\rangle$ untouched. The previously used
constant-$P$ rule is reproduced once $\beta_1=0$ or $\beta_2=0$. Fig.\
6 illustrates simulation results obtained with random homogeneous
point patterns. The relative giant component as a function of the
average transmission power $\langle{P}\rangle$ is shown for some
combinations of $\beta_1$ and $\beta_2$. All settings result in a
shift of the percolation threshold to larger $\langle{P}\rangle$
values when compared to the outcome of the constant-$P$ rule. Without
showing we remark that similar results are obtained for other point
patterns, i.e.\ multifractal or Manhattan, and other power
distributions, e.g.\ of lognormal type. The randomness and spatial
decorrelation of the long-range links introduced by the iid-$P$ link
rule does not allow for a shift of the percolation threshold towards
smaller $\langle{P}\rangle$ values. In the next Section a much more
elegant and spatially correlated approach is found to achieve this
goal.

\section{Continuum percolation with a local link rule}
\label{sec:four}

Since a central control authority does not exist in pure ad hoc
networks, there is no external provider to assign specific power
values to the ad hoc nodes according to its global rule. It is the ad
hoc network by itself which has to decide which power values are
assigned to its participating nodes. Due to the finite communication
range of each node, see again Eq.\ (\ref{eq:zweib1}), these
coordination rules have to be local. In the jargon of electrical
engineers local means distributed.  Such a local rule will be
presented and discussed in this Section.

\subsection{Local link rule: minimum degree}

By exchanging so-called hello and hello-reply messages
each ad hoc node is able to access direct information only from its
immediate neighbors, defined by its links. The simplest local
observable for a node is the number of its links, which is equal to
the number of its one-hop neighbors. Based on this observable alone, a
simple strategy for a node would be to decrease/increase its
transmission power once it has more/less than enough neighbors
\cite{RAM00}. Consequently the target node degree would be confined
between a lower and upper bound $\underline{ngb}$ and
$\overline{ngb}$; for simplicity of the following arguments we set
$\underline{ngb}=\overline{ngb}=ngb$. A value of the latter has to be
chosen such that for example almost all nodes are part of one
connected network and reflects the only external input to this
otherwise local link rule. For example, the results of the previous
Section obtained with random homogeneous point patterns suggest, that
in order to guarantee a probability almost equal to one for
one-connectivity the target value should be of the order
$O(ngb)\approx 15{-}20$; consult again Fig.\ 4 and note, that
according to the argument given in Sect.\ 3.1 in conjunction with
random homogeneously distributed point patterns and a path-loss
exponent $\alpha=2$, the relative transmission power $P/P_{\rm norm}$
can be interpreted as the average number of node neighbors.

This simple $ngb$ local link rule has at least two drawbacks. The
target range $O(ngb)\approx 15{-}20$ might be sufficient for a
randomly homogeneously patterned world of points, but for some other
underlying point patterns it would only yield a probability for
one-connectivity a little above zero. For the random multifractal
case, depicted in Fig.\ 5, the relative power has to be 
$P/P_{\rm norm}\geq 40$ for this probability to come close to one, 
which corresponds to the numerically determined value $O(ngb)\geq 52$. 
Besides this sensitivity on the specific nature of the point
patterns, also the target values $O(ngb)\approx 15{-}20$ and above are
technologically unwanted, since they lead to too much blocking for the
shared medium access control avoiding data-packet collision. The
second drawback leads to frustration due to the specific nature of the
bidirectional link definition.  A cluster of spatially close-by nodes
might saturate, meaning that each of the nodes has $ngb$ neighbors. A
new node, located not so far away from this cluster, wants to connect
to some of its nodes. In fact, the nodes of the cluster are able to
hear the lonely node due to its large transmission range, but since
they are already saturated they do not increase their power to bridge
the necessary spatial distance. The links are only one-directed, but
not bidirected. Due to the missing feedback, the lonely node further
increases its power, eventually up to its upper limit $P_{\rm max}$,
remaining in a frustrated state of having too few neighbors and
unintentionally interfering the others' communication.

In order to avoid these drawbacks we present a modified local link
rule. Upon setting up the communication links to the other nodes, a
node attaches to its hello message information about its current link
neighborhood list and its current transmission power. Starting with
$P_{\rm min}$, the node increases its transmission power by a small
amount once it has not reached a minimum link degree $ngb_{\rm min}$. 
Whenever another node, which so far does not belong to the
neighborhood list, hears the hello message of the original node for
the first time, it realizes that the latter has too few neighbors,
either sets its power equal to the transmission power of the
hello-sending node or leaves it as before, whichever is larger, and
answers the hello message.  Now the original and new node are able to
communicate back and forth and have established a new link. The
original node adds one new node to its neighborhood list. Only once
the required minimum link degree is reached, the original node stops
increasing its power for its hello transmissions. At the end each node
has at least $ngb_{\rm min}$ neighbors. Some have more because they
have been forced to answer nodes too low in $ngb$; their transmission
power is larger than necessary to obtain only $ngb_{\rm min}$
neighbors for themselves.

Fig.\ 7 illustrates the algorithmic implementation of the local
minimum-link-degree rule in more detail. Initially, all nodes come
with a minimum transmission power $P_i=P_{\rm min}$ and an empty
neighborhood list ${\mathcal N}_i=\emptyset$. All of them start in the
receive mode. Then, at random, one of the nodes switches into the
discovery mode. By subsequently sending hello messages and receiving
hello replies, the picked node increases its power until it has
discovered enough neighbors.  Then the node returns into the receive
mode. For simplicity we assume that only one node at a time is in the
discovery mode; furthermore, we assume the maximum transmission power
$P_{\rm max}$ to be sufficiently large, so that each node is able to
discover at least $ngb_{\rm min}$ neighbors.  Another node, which
enters the discovery mode at a later time, performs the same
operations. If during its previous receive-mode period this node had
already been sending hello replies to then discovering nodes, it has
to execute one more operation before returning into the receive mode:
it compares its final discovery transmission power $P_i^{\rm disc}$
with the maximum power $P_i$ it has been asked to transmit hello
replies; if the former is smaller than the latter, the node updates
its neighborhood list by sending a further hello message with the
power $max(P_i^{\rm disc},P_i)$ and receiving additional hello
replies.  -- In the receive mode a node listens to incoming hello
messages. Upon receipt of such a message, the node first checks
whether it already belongs to the incoming neighborhood list. If yes,
the requesting node has already asked before with a smaller discovery
power and there is no need for the receiving node to react. Otherwise,
it updates its transmission power to $\max(P_i,P_j)$, but only if the
magnitude of the incoming neighborhood list is smaller than the
required minimum link degree. Then it sends back a hello reply. If the
node in the receive mode has already executed its discovery mode
before, an additional operation is needed to update its neighborhood
list: it sends again a full hello message and listens to the newly
triggered hello replies. Upon doing so, the node checks on the status
of the initial hello reply, whether node $j$ was able to receive it or
not, and, if yes, makes sure that both nodes are registered in their
mutual neighborhood lists. Also some hidden links are identified by
this additional procedure; a hidden link occurs when two nodes, having
already executed their discovery mode, are forced to increase their
respective transmission powers by an independent third and fourth
party in such a way, that they are then able to communicate directly
to each other, but do not yet have this knowledge.

The implementation of the local minimum-link-degree rule written down
in Fig.\ 7a represents the ad hoc node's view, which the latter use to
selforganize into a network. The transmission power of each node is
locally chosen to adapt to the spatial surrounding and is not globally
assigned from outside.  
-- 
As simulators we employ an equivalent, but much simpler from-outside
implementation. It is depicted in Fig.\ 7b. For each node $i$ its
relative distances $r_{ij}$ to the other nodes $j$ of the point
pattern are sorted in increasing order. The first $ngb_{\rm min}$
nodes of this list make up the discovery list ${\mathcal N}^{\rm
disc}_i$. Invoking (\ref{eq:zweib1}), the discovery power 
$P_i=P^{\rm disc}_i=(r_{ij(ngb_{\rm min})})^\alpha$ is fixed by the 
relative distance belonging to the $ngb_{\rm min}$'th node of the 
sorted list, but needs to be updated according to 
$P_i=\max(P_i,(r_{ji})^\alpha)$ once the node $i$ falls into the 
discovery neighborhood of the other nodes $j$. After completion of 
the heterogeneous power assignment, the links are easily constructed 
along the lines of the definition given in Sect.\ 2.2, which then fix 
the final neighborhood lists ${\mathcal N}_i$.

\subsection{Giant component and one-connectivity}

Figs.\ 5 and 8 illustrate the simulation results obtained with the
local minimum-link-degree rule and compares them to the respective
outcomes of the artificial constant-$P$ rule. For geometric graphs
based on random homogeneous spatial point patterns the threshold of
the average giant component is reduced by about a factor of $1.45$;
for the probability for one-connectivity the threshold reduction
factor around $2.05$ is even slightly larger. Note that for 
$ngb_{\rm min}=3$ the relative giant component is already very close 
to one; once $ngb_{\rm min}\geq 6$ the probability for 
one-connectivity becomes one almost surely.

For geometric graphs based on random multifractal spatial point
patterns the local minimum-link-degree rule beats the artificial
constant $P$ rule even more impressively. Whereas for the artificial
rule the giant component threshold is very blurred, the local rule
transforms it into a sharp threshold, which almost exactly coincides
with the respective threshold obtained for the previously discussed
random homogeneous point patterns. Due to the strong spatial
clustering the artificial constant-$P$ rule leads to a highly
suppressed one-connectivity yield; for the parameters used for Fig.\ 5
the onset for nonvanishing probability is already at a rather large
transmission power around $P/P_{\rm norm}\approx 13$, but it needs an
even much larger $P$ for this probability to come close to one.  The
local rule, on the other side, acts like a wondrous outperformer,
compensating again the strong spatial clustering introduced by the
inhomogeneous spatial point patterns and pushing the one-connectivity
threshold down to $\langle P\rangle/P_{\rm norm}\approx 6$, nearly
matching the respective homogeneous-point-pattern threshold; with
$ngb_{\rm min}\approx 7$, which is equivalent to $\langle
P\rangle/P_{\rm norm}\approx 10$, the probability for one-connectivity
is practically one.

In case of the geometric graphs based on random Manhattan spatial
point patterns and in comparison with the artificial constant-$P$
rule, the local minimum-link-degree rule reduces the threshold of the
average giant component as well as of the probability for
one-connectivity by about a factor of $1.7$. The magnitude of
reduction is comparable to the values stated for random homogeneous
point patterns. For the one-connectivity probability to become almost
one, a value of at least $ngb_{\rm min}=10$ is needed for the model
parameters stated in Fig.\ 5.

\subsection{Node degree and power transmission distribution}

Next, we illustrate the node degree and transmission power distributions
resulting from the minimum-link-degree rule and compare them with their
constant-$P$ counterparts. At first random homogeneous spatial point 
patterns are considered and the minimum link-degree is chosen to be
$ngb_{\rm min}=6$ in order to guarantee one-connectivity almost surely; 
consult again Fig.\ 5. The node degree distribution $p(k)$, which
reflects the probability for a node to have $k$ neighbors, is shown in
Fig.\ 9(a1). By construction, $p(k)=0$ for $k<ngb_{\rm min}$. It comes 
with a peak at $k=ngb_{\rm min}$ and falls off sharply for 
$k>ngb_{\rm min}$. Upon changing from discrete to continuous 
$k\geq ngb_{\rm min}$, the distribution can be nicely fitted as the 
superposition $p(k) = a\delta(k-ngb_{\rm min}) + b N(k;\mu,\sigma)$ of a 
$\delta$-function, placed at the minimum link-degree, and a normalized
Gaussian with shift $\mu$ and width $\sigma$. The parameters used for the
fit in Fig.\ 9(a1) are $a=0.14$, $b=1.49$, $\mu=5.93$ and $\sigma=2.39$.
--
For comparison we also show the much broader node degree distribution 
resulting from the constant-$P$ rule. The setting $P/P_{\rm norm}=20$ 
is necessary to guarantee one-connectivity almost surely; consult again 
Fig.\ 5. This distribution can be nicely fitted with a normalized 
Gaussian; parameters used in Fig.\ 9(a1) are $\mu=18.98$ and 
$\sigma=4.92$. The small deviations to the expected Poissonian 
$p(k)=(\lambda^k/{k!})e^{-\lambda}$ with mean $\lambda=P/P_{\rm norm}$
(see Section 3.1), which is also illustrated in this Figure, are due to 
finite-size effects, that on average nodes close to the boundary 
experience a degree less than $\lambda$.

The power transmission distribution in case of the constant-$P$ rule is 
simply a $\delta$-function. It is indicated as an arrow within Fig.\ 
9(a2). Its position naturally represents an approximate upper bound for 
power transmission values obtained from the minimum-link-degree rule.
For the distribution resulting from the minimum-link-degree rule an 
analytic estimate can be given. For simplicity the path-loss exponent is 
set to $\alpha=2$. As discussed already in Sect.\ 3.1, the degree 
distribution of a node, not too close to a boundary, is given by the 
Poissonian $p(k) = (qN)^k e^{-qN}/{k!}$, with 
$q = (P/P_{\rm norm}) (\pi R_{\rm norm}^2/L^2)
   = (P/P_{\rm norm})/N $ 
representing the area covered by the node's transmission radius relative 
to the total domain area $L^2=1$. Here, the transmission power $P$ is 
kept fixed and the degree $k$ is the discrete random variable. Another
look on Fig.\ 9(a1) reveals, that the very narrow degree distribution
resulting from the minimum-link-degree rule can be crudely approximated
as $p(k)\approx\delta(k-k_0)$ with $k_0=\langle k\rangle=7.36$. Hence,
the node degree is now kept fixed within the above Poissonian
and the transmission power is considered as the continuous random 
variable. This leads to the transmission power distribution
\begin{equation}
\label{eq:vierc1}
  p(P/P_{\rm norm}) 
    \sim  (P/P_{\rm norm})^{k_0}
          \exp\left( -P/P_{\rm norm} \right)
          \; .
\end{equation}
It corresponds to a Gamma distribution 
$p(x;a,b)=x^{a-1} e^{-x/b} /(b^a\Gamma(a))$
with $x=P/P_{\rm norm}$ and $a = k_0+1 \approx 8.36$, $b=1$. The actual
best fit to the sampled distribution, shown in Fig.\ 9(a2), yields the 
parameters $a=7.74$, $b=1.01$ and more or less confirms the given 
estimate. A fit with a lognormal distribution
$p(x)=\exp\{-(\ln x-\mu)^2/2\sigma^2\}/(\sqrt{2\pi}\sigma x)$ is also 
illustrated; parameter values are $\mu=2.02$ and $\sigma=0.37$.

Node degree and transmission power distributions in connection with 
random multifractal spatial point patterns are illustrated in Figs.\
9(b1) and (b2). The minimum link-degree rule produces a node degree 
distribution, which as for the homogeneous example can be nicely 
approximated by a superposition of a $\delta$-function and a normalized
Gaussian for $k\geq ngb_{\rm min}=7$. The parameters used for the fit in 
Fig.\ 9(b1) are $a=0.13$, $b=1.36$, $\mu=7.60$ and $\sigma=3.27$. For 
comparison, the node degree distribution obtained from the constant-$P$ 
rule is also illustrated. In order to guarantee one-connectivity almost 
surely, a rather large $P/P_{\rm norm}=50$ has to be picked. This leads 
to a very broad distribution, which comes with an average degree 
$\langle k\rangle=62.7$ and which can be represented as a superposition 
$p(k)=a_1N(k;\mu_1,\sigma_1)+a_2N(k;\mu_2,\sigma_2)+a_3N(k;\mu_3,\sigma_3)$
of three normalized Gaussians with parameters 
$a_1=0.04$, $\mu_1=17.8$, $\sigma_1=5.5$,
$a_2=0.48$, $\mu_2=40.4$, $\sigma_2=13.7$,
$a_3=0.48$, $\mu_3=89.5$, $\sigma_3=22.0$.
The occurence of the double-hump structure is specific to the chosen 
parameters $N=1600$, $\beta=0.4$ and $P/P_{\rm norm}=50$, where the 
transmission range area associated to the transmission power smoothes out 
any substructure generated beyond the iteration step $j\approx 2{-}3$ of 
the multifractal point pattern construction.
--
The sampled transmission power distribution resulting from the minimum
link-degree rule is exemplified in Fig.\ 9(b2). A best fit to a Gamma
and a lognormal distribution reveals better agreement with the latter.
Best parameters are $a=2.11$, $b=3.94$, and $\mu=1.97$, $\sigma=0.81$, 
respectively.

For completeness, node degree and transmission power distributions in 
connection with random Manhattan spatial point patterns are illustrated 
in Figs.\ 9(c1) and (c2). The choice $P/P_{\rm norm}=25$ guarantees
one connectivity almost surely in context with the constant-$P$ rule and
leads to a rather broad Gaussian-like node degree distribution. The 
corresponding curve in Fig.\ 9(c1) comes with mean $\mu=28.3$ and 
width $\sigma=6.8$. Note, that the transmission radius corresponding to 
$P/P_{\rm norm}=25$ equals about half the length $1/N_x=1/N_y$ of a 
Manhattan block, so that the influence of the Manhattan structure still
impacts the node degree distribution and has not been washed out to 
coincide with a node degree distribution of a random homogeneous point 
pattern with $\mu\approx 25$. This difference also holds for the
node degree distribution resulting from the minimum link-degree rule.
The peak at $k=ngb_{\rm min}=10$ is very pronounced. The overall mean
is $\langle k \rangle = 13.53$ and the distribution 
$p(k\geq ngb_{\rm min}) \approx a\delta(k-ngb_{\rm min}) 
 + b_1N(k;\mu_1,\sigma_1) + b_2N(k;\mu_2,\sigma_2)$
can be approximated as a superposition of a $\delta$-function
and two normalized Gaussians. Parameters used in Fig.\ 9(c1)
are $a=0.35$, $b_1=0.63$, $\mu_1=7.78$, $\sigma_1=2.31$, and 
$b_2=0.56$, $\mu_2=15.90$, $\sigma_2=4.60$.
--
The sampled transmission power distribution resulting from the minimum
link-degree rule is exemplified in Fig.\ 9(c2). A best fit to a Gamma
and a lognormal distribution reveals a slightly better agreement with the 
latter, but deviations from both are clearly visible. Best parameters are 
$a=4.00$, $b=2.57$, and $\mu=2.27$, $\sigma=0.54$, respectively.

\section{Conclusion}
\label{sec:five}

For wireless mobile ad hoc communication networks connectivity
represents an important issue. In a selforganizing manner, the
participating ad hoc nodes have to tune their transmission powers to
establish direct one-hop communication links to their spatial
neighbors and to be able to reach all others via multihop routes. In
the static limit, two-dimensional continuum percolation based on discs
with constant or random radius can be mapped onto artificial link
coordination rules with constant or random transmission
power. Basically the transmission powers have to be chosen above a
percolation threshold in order to guarantee strong network
connectivity almost surely. However, the percolation threshold does
show a sensitive dependence on the specific spatial patterning of the
ad hoc nodes. Different classes of uncorrelated and correlated random
point patterns, like homogeneous, multifractal or Manhattan-like
distributions, make a big difference. The artificial link rules are
not flexible enough to adapt to local spatial inhomogeneities. A local
generalization of these rules leads to the minimum-link-degree
rule. It can be viewed as a step towards a selforganizing mutation of
continuum percolation.  It requires each ad hoc node to be connected
to a minimum number of closest neighbors, which sets a lower bound on
the node's transmission power. This node, on the other hand, might be
forced to increase its transmission power further, in order to
establish an additional communication link to another node, which
belongs to the greater vicinity and has not yet reached the required
minimum number of closest neighbors.  This distributed rule is able to
counterbalance local spatial inhomogeneities occurring in the random
point patterns. As a function of average transmission power the
percolation thresholds associated to the various considered classes of
random point patterns almost collapse onto each other and are
tremendously reduced, when compared to the outcomes with the
artificial rules.

Compared to other local link rules yielding strongly connected
networks, the presented minimum-link-degree rule is simple.  For
example, the Rodoplu \& Meng algorithm \cite{ROD99} relies on
GPS-based position knowledge of the ad hoc nodes to construct a link
enclosure for each node; the construction also requires knowledge
about the assumed uniform path-loss exponent, which can be seen as
another drawback of this algorithm. Another proposal
\cite{LIH02} requires ad hoc nodes to come with directional antennas; 
strong network connectivity is then guaranteed once each node has
neighbors in each angular sector. Rules like these require more
technological equipment for the ad hoc nodes. In terms of the
simplicity principle, the presented minimum-link-degree rule appears
to be more attractive. It also has the advantage that it will work in
a heterogeneous propagation medium, where the path-loss exponent is a
function of position, distance and direction. A generalization of the
distributed minimum-link-degree rule, which so far only has been
developed for the static limit, towards mobile ad hoc networks is
straightforward. These last two issues will be discussed in more
detail in future work.

Finding efficient distributed coordination rules, i.e.\ protocols, for
connectivity is certainly one important issue for ad hoc communication
networks, but there are definitely also others: power consumption,
efficient routing discovery and execution, medium access control,
interference, quality of service, and end-to-end throughput. The
construction of an optimized protocol for one part alone, needs not be
the overall best for all of the partially conflicting entities
considered together. Clearly, the design of a rather complex, but
still simple overall protocol is called for. It is possible that such
an optimized protocol might lead to small-world or scale-free
geometric-graph topologies, which are already observed and discussed
for Internet and biochemical communication networks
\cite{ALB02,DOR02}.

\newpage
\ack
The authors acknowledge fruitful discussions with Michael Bahr, 
Rainer Sauerwein, Clemens Hoffmann and Bernd Sch\"urmann.  
W.\ K.\ acknowledges support from the Ernst von Siemens-Scholarship.


\newpage
\begin{figure}
\begin{centering}
\epsfig{file=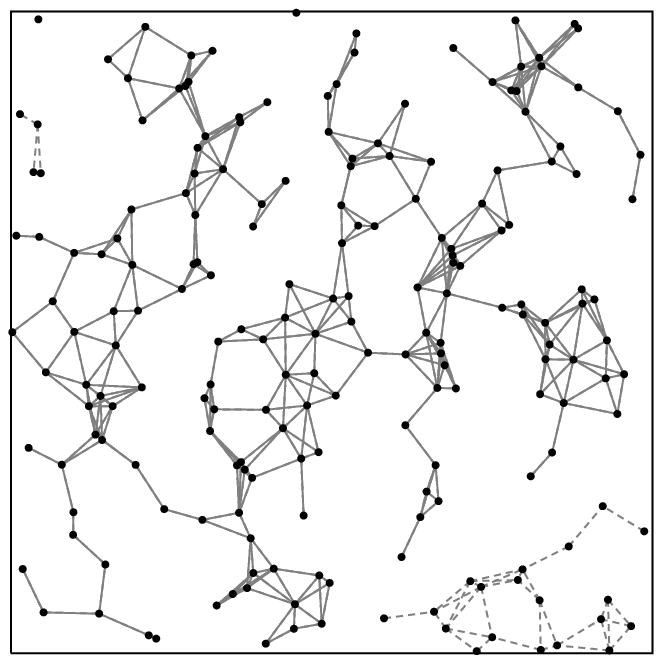,width=7cm}
\epsfig{file=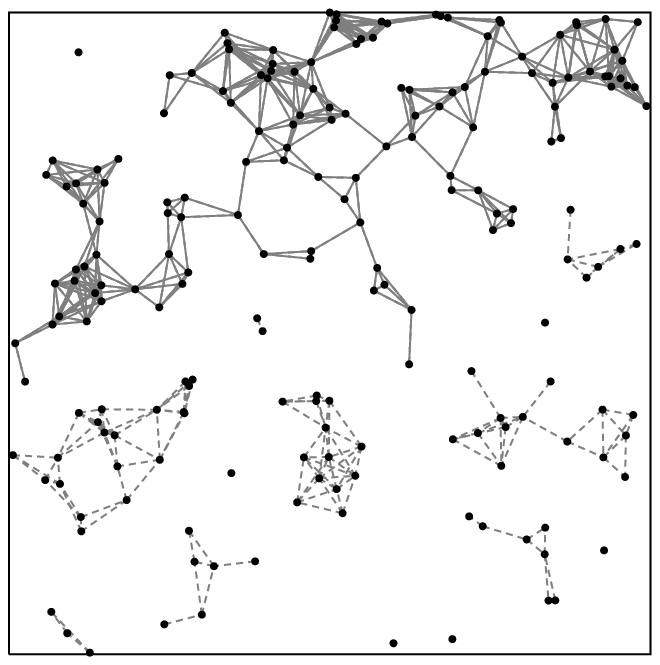,width=7cm}
\epsfig{file=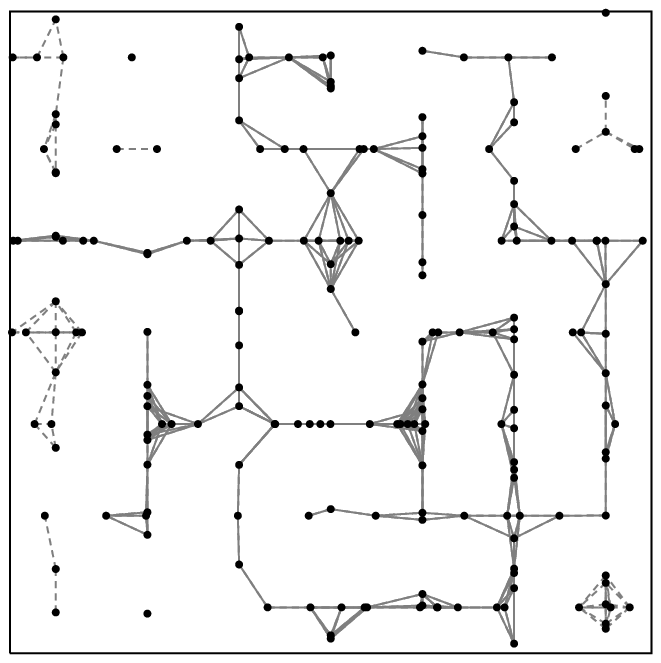,width=7cm}
\caption{
Geometric graphs for random patterns of $N=200$ points confined to a 
box: homogeneous (top), multifractal (middle) and Manhattan (bottom). 
Parameters of the point patterns are $j=5$, $\beta=0.4$ (multifractal) 
and $N_x=N_y=7$ (Manhattan). Each point is given the same transmission 
power $P/P_{\rm norm}=5$, corresponding to a link range $R/L=0.089$ 
when using the path-loss exponent $\alpha=2$. Points connected by solid 
links belong to the giant-component cluster.
} 
\end{centering}
\end{figure}

\newpage
\begin{figure}
\begin{centering}
\epsfig{file=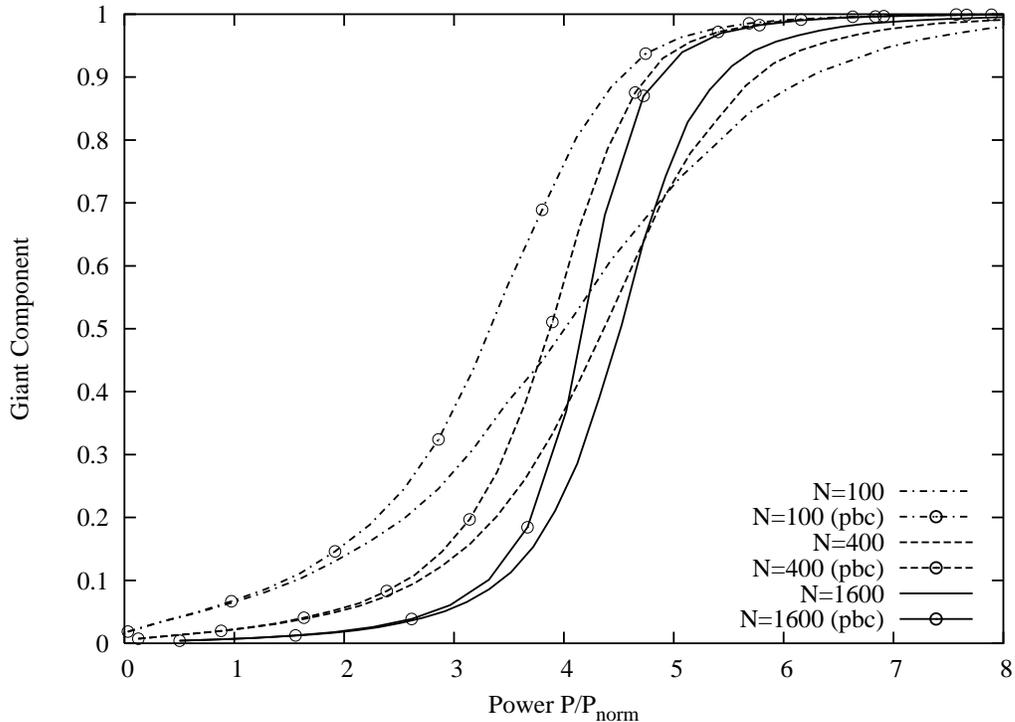,width=14cm}
\caption{
Average relative giant component as a function of transmission
power. A sample of 500 geometric graphs generated with random
homogeneous spatial point patterns and the constant-$P$ rule has been
used. The path-loss exponent of Eq.\ (\ref{eq:zweib1}) has been set to
$\alpha=2$.  Different curves correspond to different number of nodes:
$N=100$ (dash-dotted), $400$ (dashed), $1600$ (solid); curves marked
without/with open circles correspond to exclusion/inclusion of
periodic boundary conditions (pbc).  
}
\end{centering}
\end{figure}

\newpage
\begin{figure}
\begin{centering}
\epsfig{file=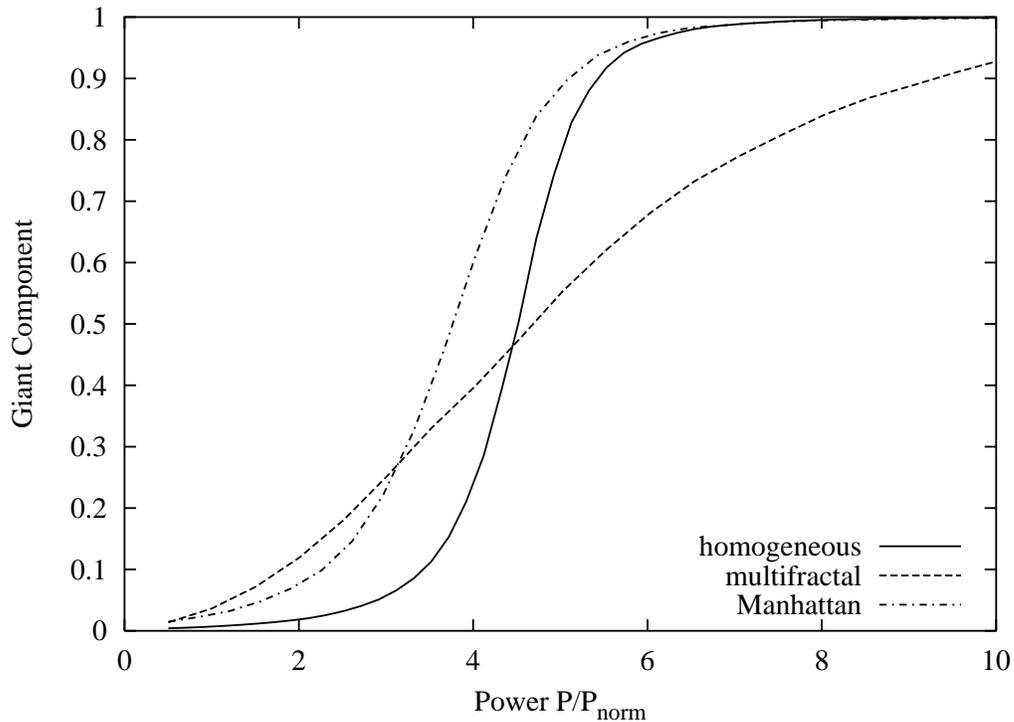,width=14cm}
\caption{
Average relative giant component as a function of transmission power 
upon using the constant-$P$ rule. The number of nodes has been fixed to 
$N=1600$ and the path-loss exponent has been set to $\alpha=2$. Different 
curves correspond to different random spatial point patterns: 
homogeneous (solid), multifractal (dashed) with parameters $\beta=0.4$ 
and $j=5$, and Manhattan (dash-dotted) with parameters $N_x=N_y=7$. A 
sample of 500 geometric graphs has been used for each case. 
} 
\end{centering}
\end{figure}

\newpage
\begin{figure}
\begin{centering}
\epsfig{file=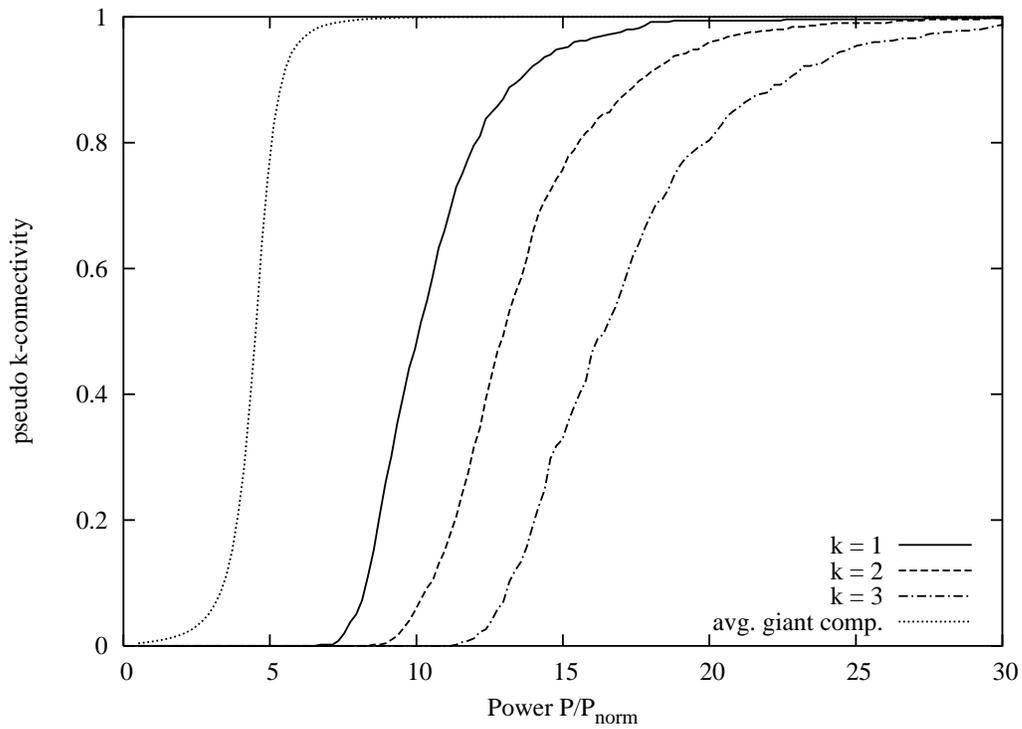,width=14cm}
\caption{
Probability of pseudo $k$-connectivity as a function of transmission power 
upon using the constant-$P$ rule. The number of nodes for the simulated 
500 random homogeneous point patterns has been fixed to $N=1600$ and the 
path-loss exponent has been set to $\alpha=2$. Different curves correspond to 
$k=1$ (solid), $2$ (dashed), $3$ (dash-dotted); for comparison the average 
relative giant component is shown as the dotted curve.
} 
\end{centering}
\end{figure}

\newpage
\begin{figure}
\begin{centering}
\epsfig{file=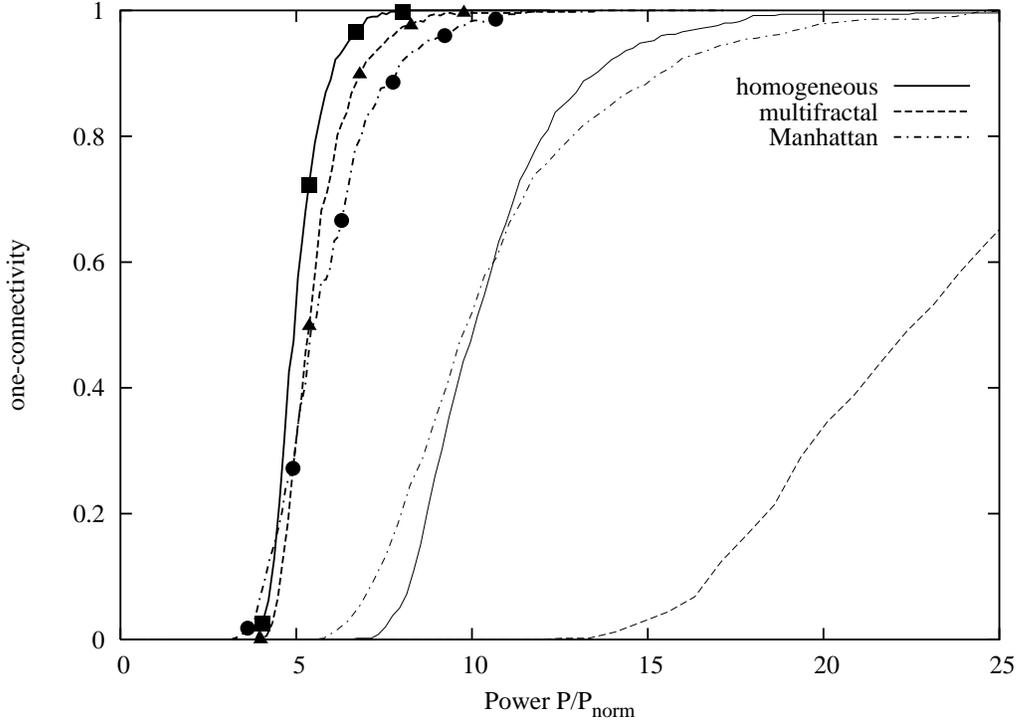,width=14cm}
\caption{
Probability for one-connectivity as a function of average transmission 
power upon using the local minimum-link-degree rule (curves with symbols) 
and the artificial constant-$P$ rule (curves without symbols). The number of 
nodes has been fixed to $N=1600$ and the path-loss exponent has 
been set to $\alpha=2$. Different line types of the curves correspond to
different random spatial point patterns: homogeneous (solid), multifractal 
(dashed) with parameters $\beta=0.4$ and $j=5$, Manhattan (dash-dotted) 
with parameters $N_x=N_y=7$; a sample of 500 geometric graphs has 
been used for each case. From the left to the right the symbols on 
each minimum-link-degree-rule curve stand for $ngb_{\rm min}=3{-}6$ 
(homogeneous), $3{-}7$ (multifractal) and $5{-}10$ (Manhattan).
} 
\end{centering}
\end{figure}

\newpage
\begin{figure}
\begin{centering}
\epsfig{file=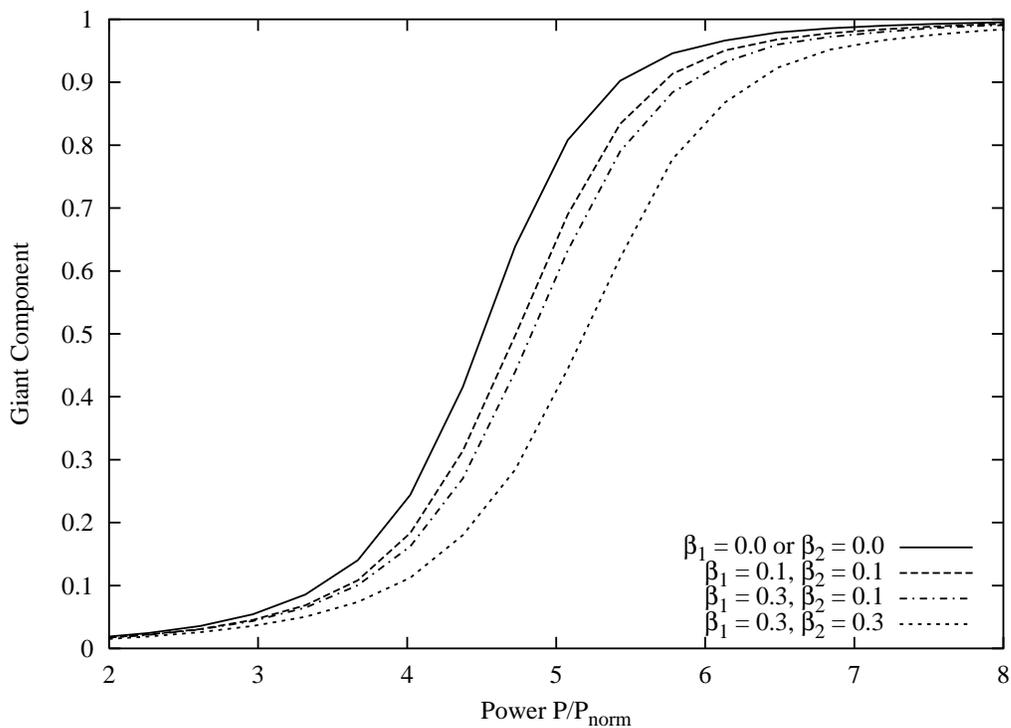,width=14cm}
\caption{
Average relative giant component as a function of the average transmission
power upon using the iid-$P$ rule. The number of nodes for each of the used 
$500$ random homogeneous point patterns has been fixed to $N=1600$ and the 
path-loss exponent has been set to $\alpha=2$. Different curves correspond 
to different parameter choices: 
$\beta_1{,}\beta_2=0.0{,}0.0$ (solid, constant-$P$ rule), $0.1{,}0.1$
(dashed), $0.3{,}0.1$ (dash-dotted), $0.3{,}0.3$ (dotted).
} 
\end{centering}
\end{figure}

\newpage
\begin{figure}
\epsfig{file=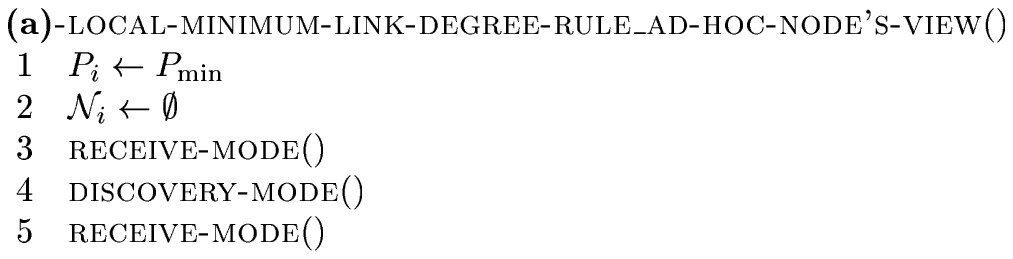,scale=1.0}
\epsfig{file=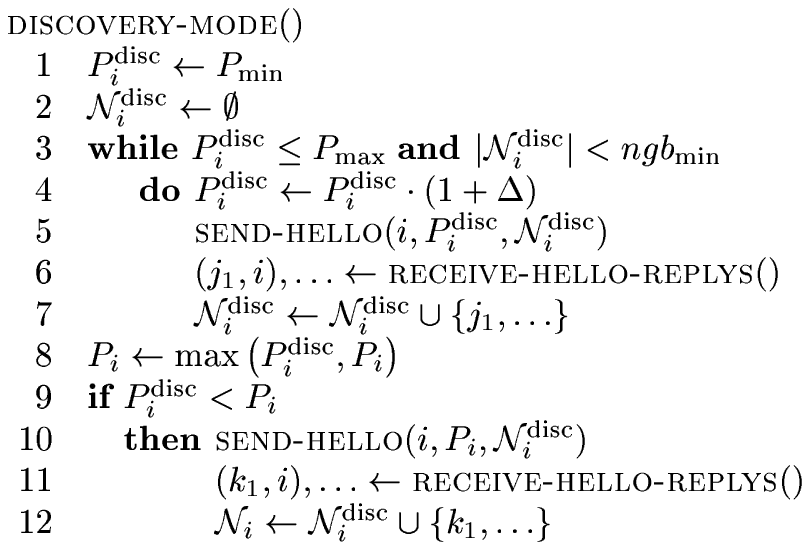,scale=1.0}\\
\epsfig{file=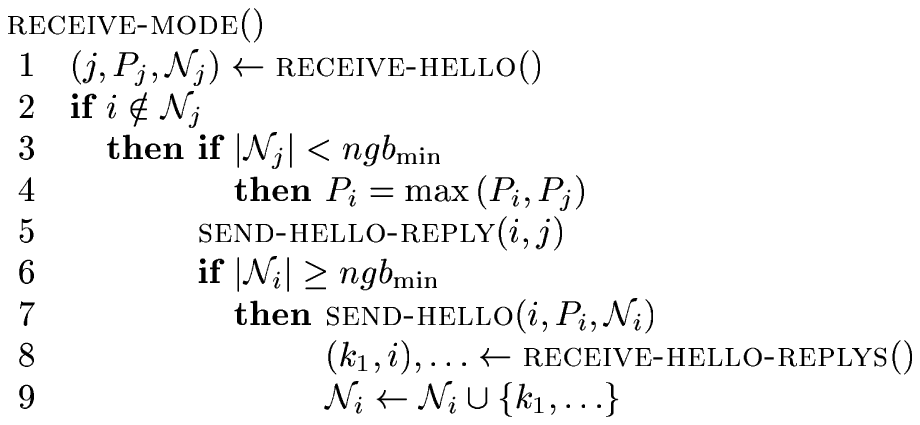,scale=1.0}\\[1cm]
\epsfig{file=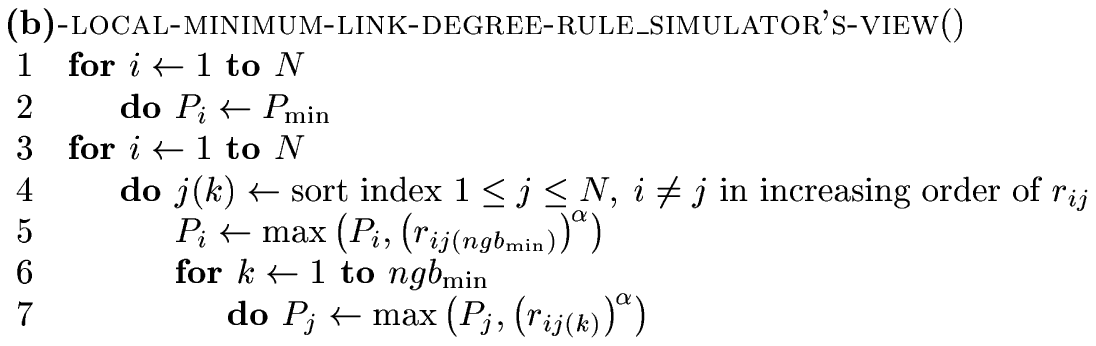,scale=1.0}
\caption{
Algorithmic implementations of the local minimum-link-degree rule:
(a) the ad hoc node's view, and (b) the simulator's view.
} 
\end{figure}

\newpage
\begin{figure}
\begin{centering}
\epsfig{file=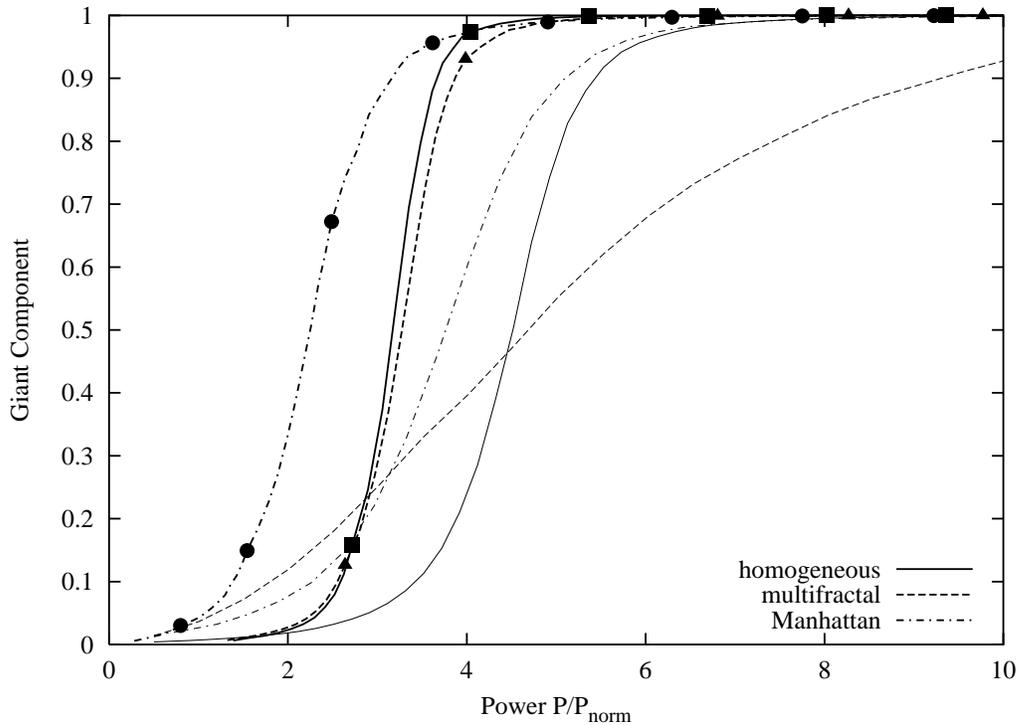,width=14cm}
\caption{
Average relative giant component as a function of average transmission
power upon using the local minimum-link-degree rule (curves with symbols) 
and the artificial constant-$P$ rule (curves without symbols, identical to 
curves of Fig.\ 3). The number of nodes has been fixed to $N=1600$ and 
the path-loss exponent has been set to $\alpha=2$. Different line types 
of the curves correspond to different random spatial point patterns: 
homogeneous (solid), multifractal (dashed) with parameters $\beta=0.4$ 
and $j=5$, Manhattan (dash-dotted) with parameters $N_x=N_y=7$; a sample 
of 500 geometric graphs has been used for each case. From the left to 
the right the symbols on each minimum-link-degree-rule curve stand for 
$ngb_{\rm min}=2{-}7$ (homogeneous, multifractal) and $2{-}9$ 
(Manhattan).
} 
\end{centering}
\end{figure}

\newpage
\begin{figure}
\begin{centering}
\epsfig{file=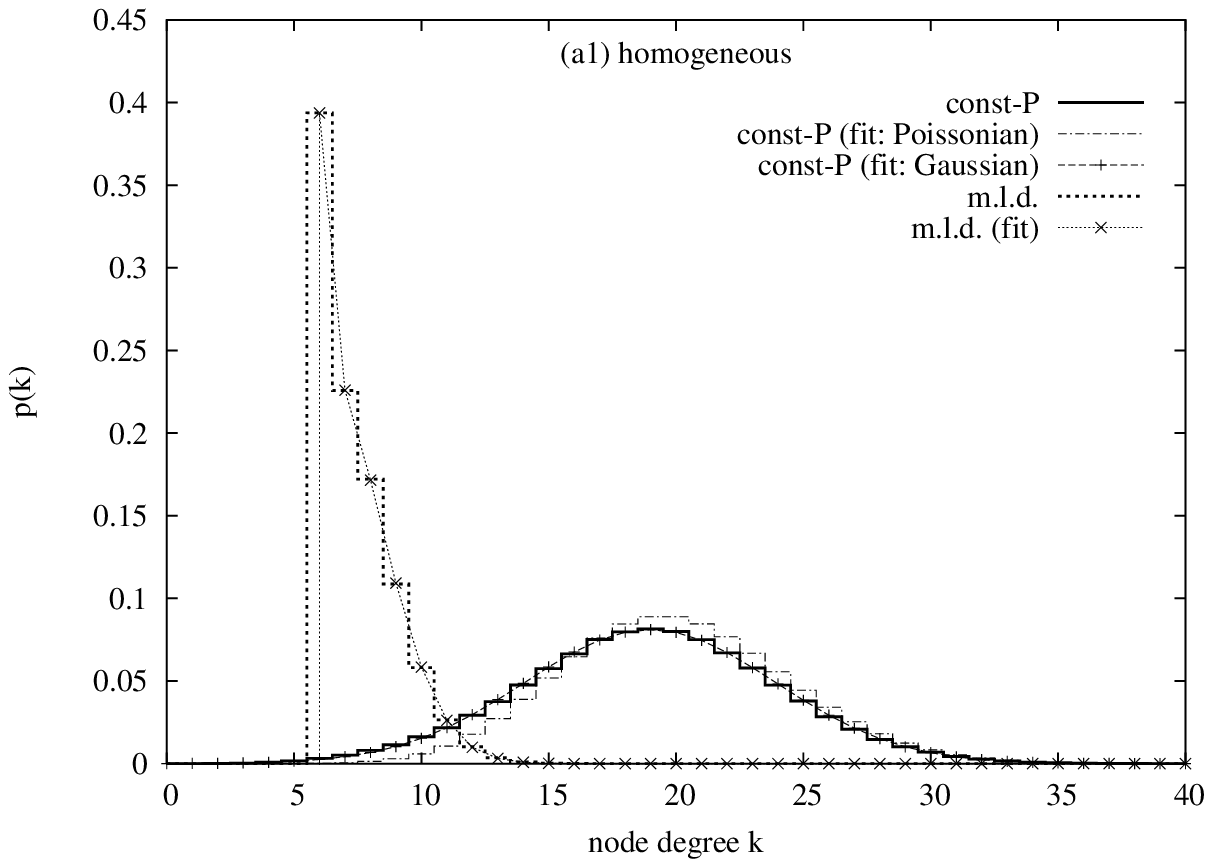,width=12cm}
\epsfig{file=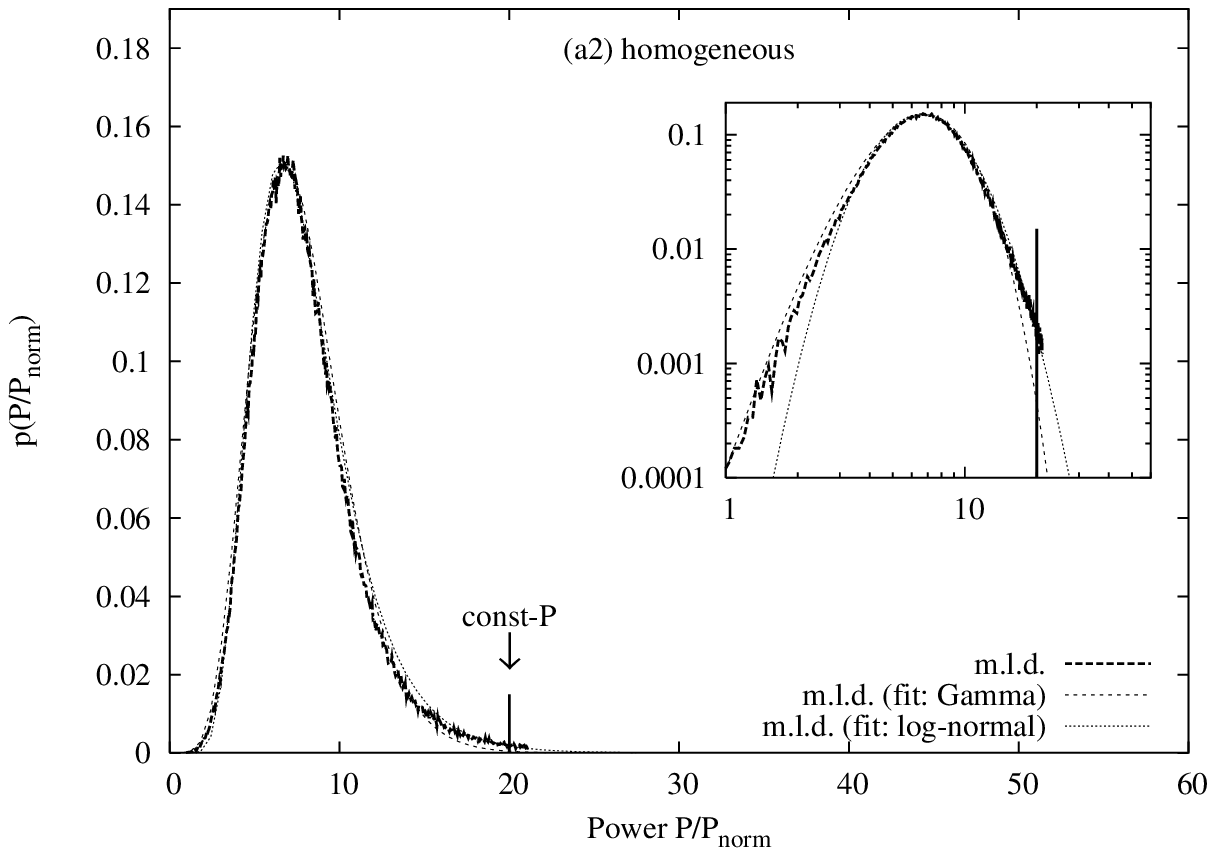,width=12cm}
\end{centering}
\end{figure}
\begin{figure}
\begin{centering}
\epsfig{file=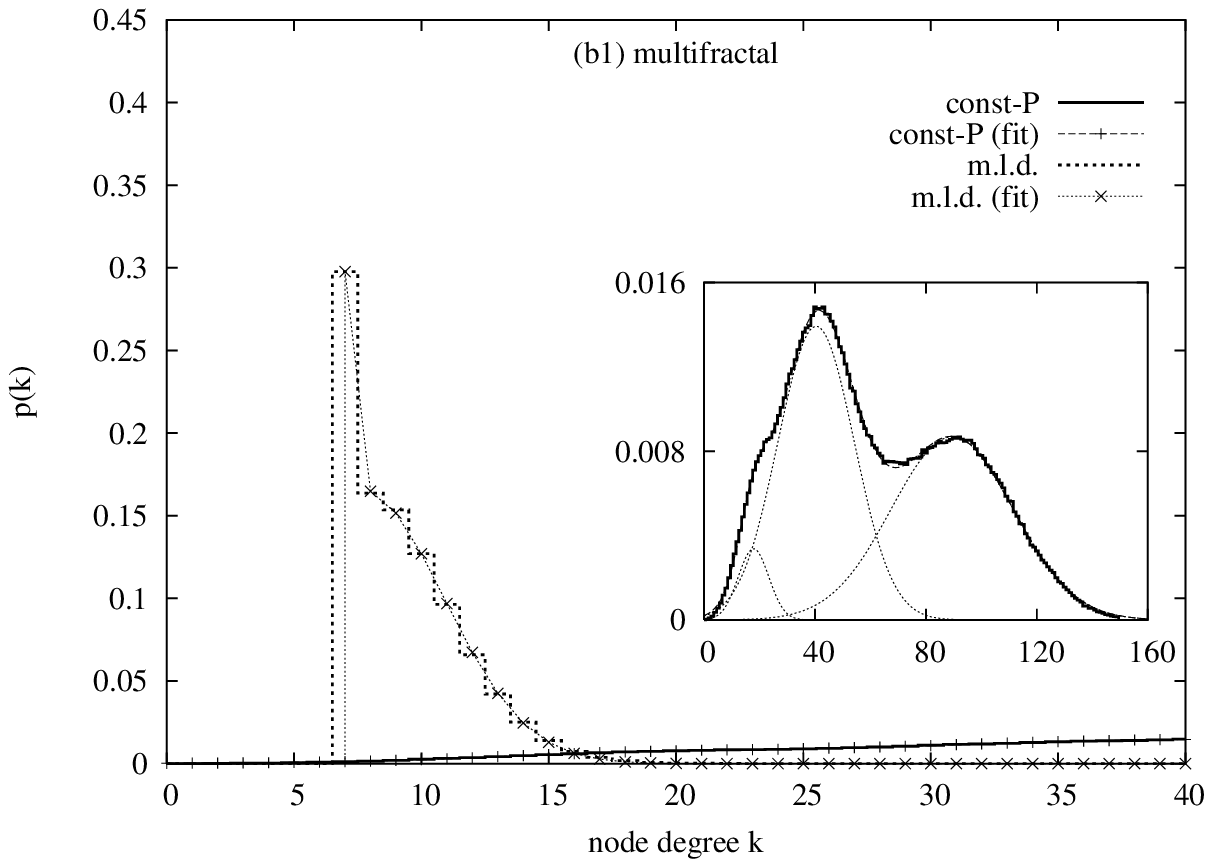,width=12cm}
\epsfig{file=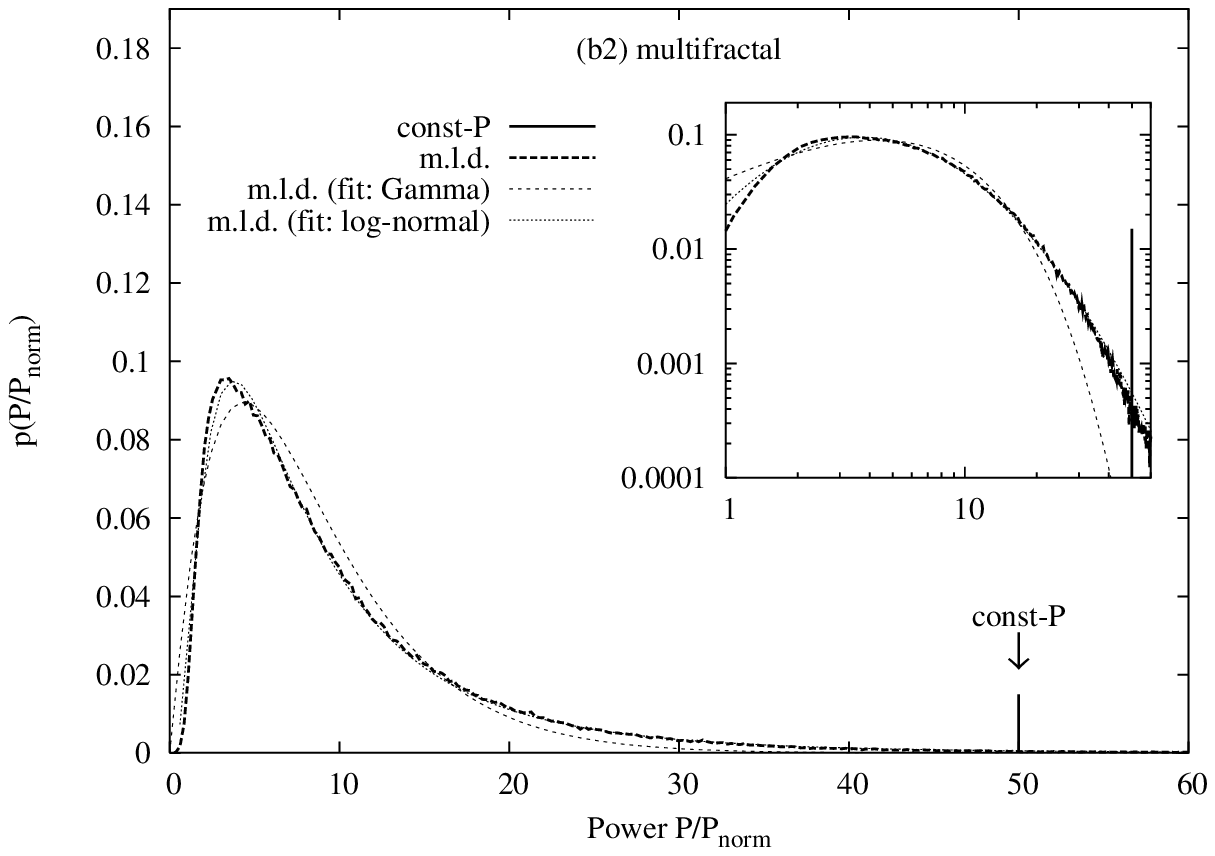,width=12cm}
\end{centering}
\end{figure}
\begin{figure}
\begin{centering}
\epsfig{file=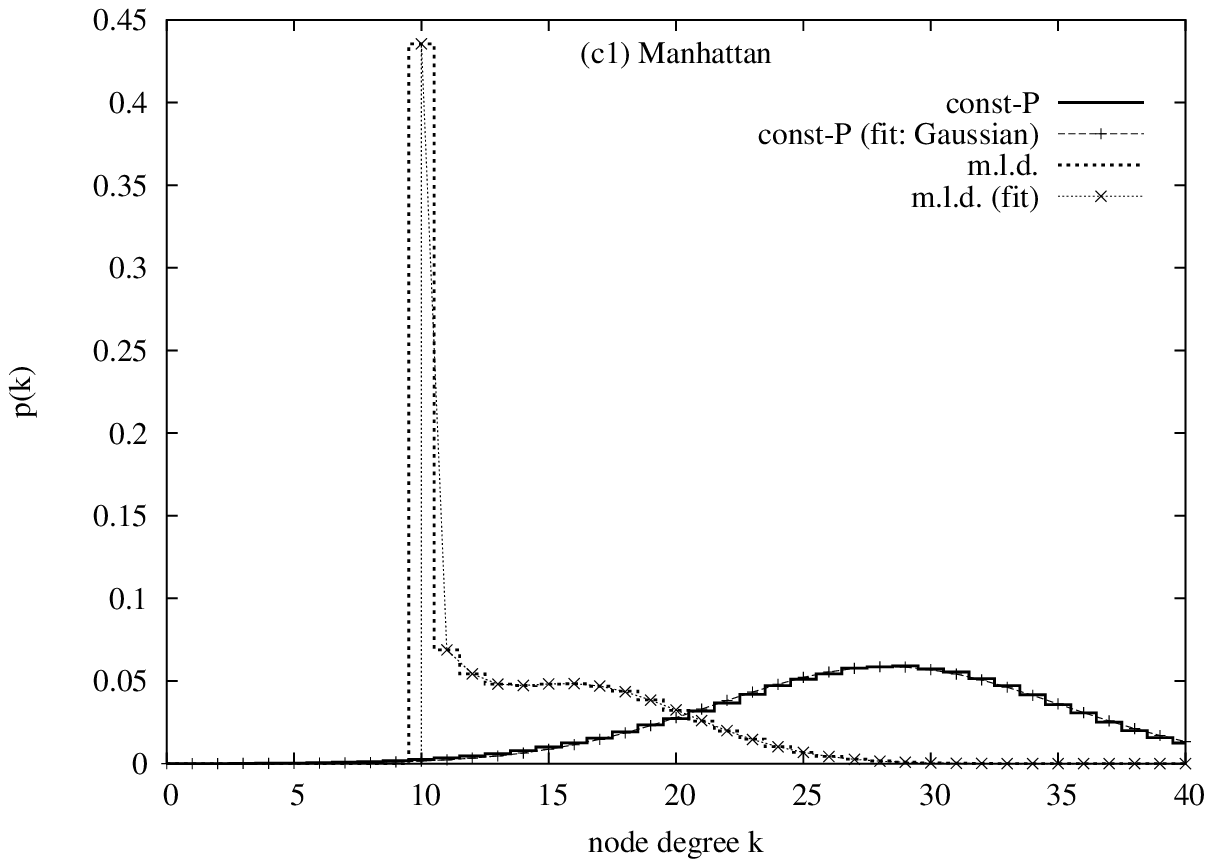,width=12cm}
\epsfig{file=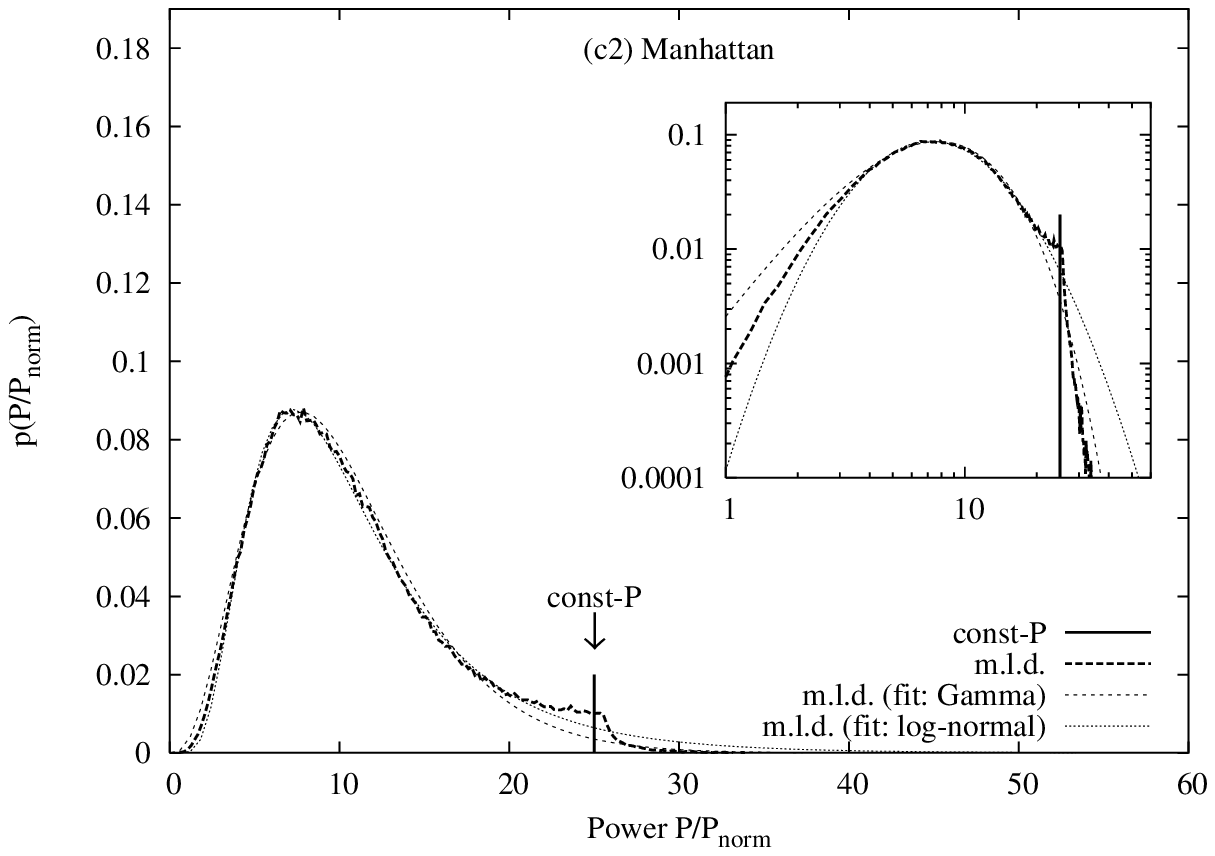,width=12cm}
\caption{
Node degree and transmission power distributions obtained from the
minimum link-degree (m.l.d.) and the constant-$P$ rule. Details about
the fitted distributions are given in the main text. The number of 
nodes has been fixed to $N=1600$ and the path-loss exponent has 
been set to $\alpha=2$. Different random spatial point patterns are used: 
homogeneous (a), multifractal (b) with parameters $\beta=0.4$ and $j=5$, 
and Manhattan (c) with parameters $N_x=N_y=7$; a sample of 500 geometric 
graphs has been used for each case.
} 
\end{centering}
\end{figure}


\begin{thebibliography}{00}




\bibitem{PRO95}
  J.G.\ Proakis, 
  {\em Digital Communications},
  McGraw-Hill, Singapore (1995).
\bibitem{PRA98}
  R.\ Prasad,
  {\em Universal Wireless Personal Communications},
  Artech House, Boston (1998).
\bibitem{RAP99}
  T.S.\ Rappaport, 
  {\em Wireless Communications -- Principles \& Practice},
  Prentice Hall, Upper Saddle River (1999).
\bibitem{HAA99}
  Z.J.\ Haas, et al., 
  eds., Special Issue on Wireless Ad Hoc Networks, 
  IEEE J. on Selected Areas in Communications, 
  Vol.\ 17, No.\ 8 (August 1999).
\bibitem{HUB01}
  J.P.\ Hubaux, T.\ Gross, J.Y.\ Le Boudec and M.\ Vetterli,
  {\em Toward Self-Organized Mobile Ad Hoc Networks: The Terminodes 
  Project},
  in IEEE Communications Magazine (Jan.\ 2001).
\bibitem{MANET}
  Mobile Ad Hoc Networks (manet) Working Group, \\
  http://www.ietf.org/html.charters/manet-charter.html.
\bibitem{NISTB}
  Wireless Ad Hoc Networks Bibliography, \\
  http://w3.antd.nist.gov/wctg/manet/manet{\_}bibliog.html.
\bibitem{MEE96}
  R.\ Meester and R.\ Roy,
  {\em Continuum Percolation}, 
  Cambridge University Press (1996).
\bibitem{STO95}
  D.\ Stoyan, W.S.\ Kendall and J.\ Mecke,
  {\em Stochastic Geometry and its Applications}, 
  John Wiley \& Sons, Chichester (1995).
\bibitem{DAL02}
  J.\ Dall and M.\ Christensen, 
  arXiv:cond-mat/0203026.
\bibitem{BAK02}
  D.R.\ Baker, G.\ Paul, S.\ Sreenivasan and H.E.\ Stanley, 
  arXiv:cond-mat/0203235.
\bibitem{ISI92}
  M.B.\ Isichenko,
  Rev.\ Mod.\ Phys.\ 64 (1992) 961.
\bibitem{BLA02}
  P.\ Blanchard, G.\ Dell'Antonio, D.\ Gandolfo and M.\ Sirugue-Collin,
  J.\ Stat.\ Phys.\ 106 (2002) 1.
\bibitem{FED88}
  J.\ Feder, 
  {\em Fractals}, Plenum Press, New York (1988).
\bibitem{FRI95}
  U.\ Frisch, 
  {\em Turbulence}, 
  Cambridge University Press (1995).
\bibitem{MEN91}
  C.\ Meneveau and K.R.\ Sreenivasan,
  J.\ Fluid Mech.\ 224 (1991) 429.
\bibitem{GRE96}
  M.\ Greiner, J.\ Giesemann, P.\ Lipa and P.\ Carruthers, 
  Z.\ Phys.\ C 69 (1996) 305.
\bibitem{GRE98}
  M.\ Greiner, H.\ Eggers and P.\ Lipa, 
  Phys.\ Rev.\ Lett.\ 80 (1998) 5333.
\bibitem{LUX01}
  T.\ Lux,
  Quantitative Finance 1 (2001) 632.
\bibitem{MUZ00}
  J.F.\ Muzy, J.\ Delour and E.\ Bacry,
  Eur.\ Phys.\ J.\ 17 (2000) 537.
\bibitem{RIB01}
  V.J.\ Ribeiro, R.H.\ Riedi and R.G.\ Baraniuk,
  {\em Wavelets and Multifractals for Network Traffic Modeling and
  Inference},
  in Proc.\ ICASSP, Salt Lake City, Utah (May 7-11, 2001).
\bibitem{WOL96}
  E.A.\ De Wolf, I.M.\ Dremin and W.\ Kittel, 
  Phys.\ Rep.\ 270 (1996) 1.
\bibitem{HAL86}
  T.\ Halsey, M.\ Jensen, L.\ Kadanoff, I.\ Procaccia and B.\ Shraiman,
  Phys.\ Rev.\ A 33 (1986) 1141.
\bibitem{BOL85}
  B.\ Bollob\'as, 
  {\em Random Graphs}, Academic Press, London (1985).
\bibitem{WHI01}
  D.R.\ White and M.E.J.\ Newman, 
  Santa Fe Institute Working Paper 01-07-035.
\bibitem{PEN99}
  M.D.\ Penrose, 
  Random Structures and Algorithms 15 (1999) 145.
\bibitem{GUP00}
  P.\ Gupta and P.R.\ Kumar, 
  IEEE Trans.\ Inf.\ Theory IT-46 (2000) 388.
\bibitem{DOU02}
  O.\ Dousse, P.\ Thiran and M.\ Hasler,
  presented at IEEE INFOCOM 2002, New York (June 23-27, 2002), 
  http://www.ieee-infocom.org/2002/papers/481.pdf.
\bibitem{STA92}
  D.\ Stauffer and A.\ Aharony,
  {\em Introduction to Percolation Theory}, 
  Taylor \& Francis, London (1992).
\bibitem{MAR02}
  M.K.\ Marina and S.R.\ Das,
  {\em Routing Performance in the Presence of Unidirectional Links in
  Multihop Wireless Networks},
  in Proc.\ 3rd ACM Int.\ Symposium on Mobile Ad Hoc Networking and
  Computing (MOBIHOC 2002), Lausanne, Switzerland (June 9-11, 2002),
  pp.\ 12-23.
\bibitem{RAM00}
  R.\ Ramanathan and R.\ Rosales-Hain,
  {\em Topology Control of Multihop Wireless Networks using Transmit
  Power Adjustment},
  in Proc.\ IEEE INFOCOM 2000 (March 2000).
\bibitem{ROD99}
  V.\ Rodoplu and T.H.\ Meng,
  IEEE Journal on Selected Areas in Communications -- 
  Special Issue on Ad Hoc Networks 17 (1999) 1333. 
\bibitem{LIH02}
  L.\ Li, J.\ Halpern, P.\ Bahl, Y.\ Wang and R.\ Wattenhofer,
  arXiv:cs.NI/0209012.
\bibitem{ALB02}
  R.\ Albert and A.L.\ Barabasi,
  Rev.\ Mod.\ Phys.\ 74 (2002) 47.
\bibitem{DOR02}
  S.N.\ Dorogovtsev and J.F.F.\ Mendes,
  Advances in Physics 51 (2002) 1079.

\end{thebibliography}
\end{document}